\documentstyle[prl,twocolumn,epsf,aps]{revtex}

\def\b{\begin{equation}}
\def\e{\end{equation}}

\begin{document}
%
%
\newcommand{\msr}{$\mu$SR}


\twocolumn[\hsize\textwidth\columnwidth\hsize\csname@twocolumnfalse\endcsname

\title{Cyclotron Emission and Thermalization of the CMB Spectrum }

\author{Niayesh Afshordi}
\address{Princeton University Observatory, Princeton, NJ 08544-1001, USA\\
 afshordi@astro.princeton.edu}

\date{\today}
\maketitle

\begin{abstract}
 
 If there was a weak magnetic field in the early universe, 
 cyclotron emission could play an important role
 in the thermalization of the CMB. We study this process 
 in the tightly coupled primordial electron-photon plasma
 and find that if the magnetic field is large enough so that  
 the plasma effects allow emission of cyclotron photons, 
 this process will wipe out deviations from the black body
 spectrum.
   
\vspace{5mm}
\end{abstract}
\vspace{7mm}
]


 The observed spectrum of the Cosmic Microwave Background
 radiation is beautifully fit by a black body
 spectrum. The FIRAS instrument on the COBE 
 satellite measured this spectrum and found that deviations from
 a Planck spectrum 
 are less than a few hundredth of a percent. In particular, 
 the chemical potential potential is less than $9 \times 10^{-5}$
 \cite{fixsen}. This puts strong
 constraints on any non-thermal photon production
 mechanism, for example, the decay \cite{burigana},\cite{hu}, or 
 annihilation \cite{mcdonald}
 of relic particles after $z \sim 10^7$. After $ z \sim 10^7$, the
 previously studied thermalization processes, i.e. bremsstrahlung 
 and double Compton,
  can no longer significantly change 
 the number of photons. As a lower limit, a value of
 $\sim 10^{-9}$ is expected for the chemical potential,  
 in the absence of these processes. This a consequence of 
 dissipation of the acoustic waves at small scales  
 prior to the recombination\cite{peebles}.      

  There are indications that a primordial 
 magnetic field is necessary to explain the observed magnetic
 fields in galaxies and galaxy clusters 
 (see \cite{grasso} and references therein
 for a review of the issues relevant to the primordial magnetic 
 field). In the absence of any dynamo effect, the required field 
 strength today is of the order of $10^{-9} G$ in the Mpc scales.

 There are many proposed mechanisms for generating magnetic
 fields. They could be produced through QCD (e.g. \cite{david},\cite{cheng}) and Electroweak 
 \cite{baym} phase transitions. The magnitude of the generated field 
 depends on scale and the details of the phase transition model. Although
 the typical predictions are several orders of magnitude less than the 
 required field strength ($\sim 10^{-20} G$, e.g. \cite{sigl}) on the
 relevant scales, 
 there are extreme 
 models that can produce large enough field strengths\cite{grasso}.
 However, note that the generated field can be significantly larger 
 at small scales.

  There are two main stages in the thermalization of the 
 radiation field. The first stage is relaxation into a 
 Bose-Einstein energy distribution which is mainly through
 Compton scattering of photons off electrons. The number of 
 photons is fixed during this stage. This process, so-called
 Comptonization, is efficient for temperatures higher than $10 
 \, {\rm eV}$. The second stage is relaxation of the chemical
 potential which, in general, requires an increase in the number 
 of photons. The most efficient processes in the early universe 
 that can change the number of photons are bremsstrahlung and double
 Compton. In a universe with low barion density 
 , double Compton is more important than bremsstrahlung but 
 ceases to be efficient for $T< 1 \, {\rm keV}$\cite{hu}.    
  
  We show that a weak magnetic field can change this 
 picture for $T< 1 \, {\rm keV}$. We see that if the 
 cyclotron frequency is larger than the plasma frequency, cyclotron
 emission thermalizes the radiation spectrum. Finally we consider the
 observational and theoretical constraints on the possiblity of
 this effect and see that plausible levels of magnetic field 
 strength may allow
 efficient cyclotron emission. 

 This phenomenon has been first considered
 in \cite{puy}. However, since the cyclotron absorption process
 was neglected, the phenomenon was interpreted as a process for
 generation rather than relaxtion of chemical potential. This
 problem was correctly pointed out in \cite{jedamzik}.

  The rate of energy loss via cyclotron emission by
 non-relativistic electrons moving in magnetic field $B$
 can be obtained classically \cite{jackson}
\b
 \frac{d{\cal E}}{dt} = \frac{2}{3}\frac{e^2\omega^2_c v^2_{\perp}}{c^3},
\e
 where $\omega_c=eB/(m_e c)$ is the cyclotron frequency and $v_{\perp}$
 is the velocity of the electron, normal to the magnetic field direction.

 In the non-relativistic limit, almost all the emitted photons
 have the frequency $\omega_c$ and thus, the rate of photon production
 per unit volume, $\phi$, can be obtained using Eq. (1)
\b
 \phi  = \frac{2}{3}\frac{n_e e^3 B < v^2_{\perp} >}{\hbar m_e c^4} = \frac{4}{3}\frac{n_e e^3 B k_BT}{\hbar m^2_e c^4},
\e
 where $n_e$ is the electron number density and $T$ is the temperature 
 of the electron gas.

 In the rest of the paper, we are going to set $k_B=\hbar=c=1$.

  The presence of photons in the environment can enhance the photon
 production mechanism through stimulated emission. Also photons can
 get absorbed by the rotating electrons. These processes can be expressed
 via \cite{spitzer}
\b
 \dot{n}(E_c) = \sum _{\{E\}}{\cal A} (1+n(E_c)) n_e(E+E_c)-
{\cal B} n(E_c) n_e(E).
\e
 Here, $n(E)$ is the photon occupation number, $E_c = \omega_c$
 is the energy of the cyclotron photons and ${\cal A}$ \& ${\cal B}$
 are Einstein coefficients. The sum, is over the energy
 states of the of electrons (Landau levels, in this case).

 We note that, since for a Planckian distribution
\b
n_{Pl}(E) = \frac{1}{\exp(E/T)-1}, 
\e
  $\dot{n}$ must vanish, (if the electrons have the same temperature
 as photons, which is the case before matter-radiation decoupling
 , when Compton scattering is efficient) we must have
\b
 \sum _{\{E\}}\exp(E_c/T){\cal A} n_e(E+E_c) = \sum _{\{E\}}{\cal B} n_e(E).
\e

  Since we are assuming that Comptonization is efficient, the only
 free parameters in the spectrum of photons are the temperature, $T$,
 and the chemical potential, $\mu$. Alternatively, they can be
 replaced by the total number density, $N$,  and the energy density $u$,
 of the photon gas. As a result, we can integrate Eq. (3) over the phase
 space to obtain the total photon injection rate, without losing any
 information. Multiplying this by $E_c$, gives the energy injection rate.
\b
 \dot{N} = 2\int d^3p  \sum _{\{E\}}{\cal A} n_e(E+E_c) (1+n(E_c)-
 \exp(E_c/T)n(E_c)).
\e

  The factor in front of the bracket is the photon production rate
 for zero photon occupation number, which is the same as $\phi$ in Eq.(2).
 Plugging in a Bose-Einstein energy distribution for $n(E_c)$, and
 assuming $\mu, E_c \ll T$, we end up
 with
\b
 \dot{N} = -\frac{\phi \mu }{E_c - \mu},
\e
 and the energy production rate per volume
\b
 \dot{u} = E_c \dot{N} = -\frac{\phi E_c \mu }{E_c - \mu}.
\e

  For $\mu \ll T$, energy and number density of photons
 can be written as
\begin{eqnarray}
 u \simeq \frac{T^4}{\pi^2}[\frac{\pi^4}{15} + 7.212 \frac{\mu}{T}],
\nonumber \\
 N \simeq \frac{T^4}{\pi^2}[2.404 +\frac{2\pi^2}{3}\frac{\mu}{T}].
\end{eqnarray}

 To study the relaxation process, we replace $T$ by $T+ \theta$ and
 consider $\theta \ll T$ as the time dependent perturbation of the
 temperature, while $T$ is constant. To the first order in $\theta$ and
 $\mu$, we find
\b
 \delta u = \pi^{-2} T^3 [4\alpha \theta +\beta \mu], \, \, \delta
 N = \pi^{-2} T^2 [3\gamma \theta + \delta \, \mu ]
\e
 where
\b
 \alpha = \pi^4/15,\,\, \beta \simeq 7.212, \,\,\gamma \simeq 2.404,
 \,\, \delta= 2\pi^2/3,
\e
 are the numerical coefficients in Eq. (9).

  Now, we can use Eqs. (7)\& (8) to get the time derivatives of $\delta N$ and $\delta u$ in
  Eq. (10). We are interested in the relaxation of $\mu$, for which we
  find
\b
 \dot{\mu} = -[\frac{4\alpha \pi^2}{4\alpha \delta-3\beta
 \gamma}]\frac{\phi \mu}{T^2(E_c-\mu)}.
\e
 Note that, to the first order, $\theta$ does not appear in Eq. (12). 
 The reason is that,
 in the absence of cyclotron emission, $\mu$ and $\theta$, both remain
 constant. Also, $\mu \leq 0$ to have a finite photon occupation number 
 at all energies.
 Eq. (12) is simplified by introducing the following
 variables
\b
 \tilde{\mu} = -\frac{\mu}{E_c} , \,\, t_c = T^2 E_c \phi^{-1}[\frac{4\alpha \pi^2}{4\alpha \delta-3\beta
 \gamma}]^{-1}.
\e
 In terms of these variables, Eq. (12) takes the form
\b
 \dot{\tilde{\mu}} = -\frac{\tilde{\mu}}{t_c (1+\mu)},
\e
 with the solution
\b
 \tilde{\mu}+\ln \tilde{\mu} = -t/t_c + {\rm const}.
\e

  We see that, for $\tilde{\mu} < 1$ ($|\mu| < E_c$), there is an exponential decay with
 the characteristic time $t_c$. In the radiation dominated era, in
 a cosmological scenario, the value $t_c$ is \footnote{Assuming $T_{{\rm CMB}}
 = 2.73 {\rm K}, h_{100}= 0.7, g_{{\rm eff}} = 3.36$ and $\Omega_b =0.04$.}
\b
t_c = 3.346 (\frac{m_e T^3}{ \alpha_e n_e m_{Pl}}) H^{-1} = 1.723
\times 10^{-10} H^{-1}.
\e
  where, $m_{Pl} = G^{-1/2} $ and $\alpha_e =e^2$ are the Planck's mass and
  the fine structure constant. It is amazing
  that the ratio of $t_c$ to the Hubble time is so small and also
  independent of time and the magnetic field strength. However,
  this is not un-physical, since as $B$ goes to zero, the cyclotron
  energy, $E_c$,
  becomes smaller than $\mu$ and we are not in this regime any more.
   Since stimulated emission is proportional to the occupation
  number which goes as $E_c^{-1}$ at small energies, it cancels
  the $E_c$ that appears in $\phi$ in Eq. (2), thus the equilibrium 
  time is independent of the magnetic field strength.

  More interesting is the case of $\tilde{\mu}
 >1$ ($|\mu| > E_c$). In this case, $\dot{\tilde{\mu}} \simeq t^{-1}_c$
 is almost constant and so the relaxation time is proportional to
 the initial value of $\tilde{\mu}$.

\begin{eqnarray}
 t_{rel} = (\frac{-\mu}{E_c}) t_c = 2.22 (H
 t_c)(\frac{-\mu}{T})\beta^{-1/2}_M (\frac{T}{m_e})^{-1} H^{-1}
\nonumber\\
 = ( 6.204 \times 10^{-2} H^{-1})
 (\frac{-\mu}{T})(\frac{\beta_M}{10^{-5}})^{-1/2} T^{-1}(eV),
\end{eqnarray}
 where
\b
 \beta_M \equiv \frac{B^2}{8\pi \rho},
\e
 and $\rho$ is the energy density of the universe and so
 $\beta_M$ is the fraction of the total energy density in 
 the magnetic field (Note that $\beta_M$ remains constant 
 in the radiation dominated era). 
 At the end of this time, the chemical
 potential decays in a time scale, much shorter than the Hubble
 time.

  Since the decay is linear, rather than exponential,
  this relaxation process behaves differently from other
  processes (specifically bremsstrahlung and double Compton): 
  either $\mu/T$ is basically constant (if
  $Ht_{rel} >1$) or it is completely suppressed ($Ht_{rel}<1 $).
  Eq. (18) implies that even a dynamically negligible primordial 
  magnetic field ($B^2 \sim 10^{-6} \rho$ which is $\sim 10^{-9} G$ 
  today, if the flux  
  remains frozen) 
  can completely 
  suppress even a significant
  deviation from the Planck spectrum in the radiation dominated era.

   The natural frequency of the primordial plasma sets a lower bound
  for the frequency of propagating photons. As a result, no photon
  can be emitted via cyclotron emission if $\omega_c$ falls below
  the plasma frequency, $\omega_p = \sqrt{4\pi n_e e^2/m_e}$. Therefore this 
 thermalization
  process works only if 
 \begin{eqnarray}
  \frac{\omega^2_p}{\omega^2_c} = \frac{4\pi n_e m_e}{B^2} = 
 0.452 \, \beta^{-1}_{M} (\frac{n_e}{T^3})(\frac{m_e}{T})
\nonumber\\
 = 2.575 \times 10^{-5}\,  \beta^{-1}_M \,  T^{-1}({\rm eV}) \, < \, 1,
\end{eqnarray}
 which yields
\b
 \beta_M = \frac{B^2}{8 \pi \rho} >  2.575 \times 10^{-5} T^{-1}({\rm eV}).
\e
 or in terms of the magnetic field strength at the present time
\b
 B > 2.3 \times 10^{-8} T^{-1/2}({\rm eV}) {\rm G} 
\e      

  It is reasonable to assume that the maximum value of 
 $-\mu/T$, as a result of a non-thermal process, is close to one,
 since $\mu$ and $T$ are affected in similar ways by non-thermal
 processes (see Eq.(10)). With this assumption, combining  Eqs. (17)
 \& (20) yields
\b
 H t_{rel} < 3.866 \times 10^{-2} T^{-1/2}({\rm eV}) <1, 
\e
  for $T > 10^{-3} {\rm eV}$, which is clearly the case in the radiation 
 dominated era. This leads to the main conclusion of this letter: 
 {\it if the condition of Eq. (20)
 is satisfied, any deviation from the Planck spectrum will be completely    
 suppressed.} This will be basically indpendent of the origin or the 
 magnitude of this deviation. 
 
  The strongest observational constraint on the magnitude 
 of a primordial magnetic field comes from the spectrum of 
 the CMB fluctuations \cite{Mack}. The current upper limit
 on $\beta_M$ is about $10^{-10}$ at the comoving scale of 
 $\sim 1 \, {\rm Mpc}$. However, the actual strength of the 
 magnetic field can be significantly larger at smaller scales.
 For example, with the random dipole approximation \cite{grasso},
 $\beta_{M} \propto r^{-3} $ and so we see that the observational 
 constraint hardly gives us any information about the possibility 
 of cyclotron thermalization process if the field extends down
 to the scale of $r \sim 10 \, {\rm kpc}$. If the actual magnitude
 of the field is anywhere close to the observational limit at this 
 scale, it is likely that $\beta_M$ will satisfy Eq.(20) at smaller scales
 and the cyclotron emission will suppress any chemical potential
 in the CMB.  

  A direct method has been suggested in \cite{jedamzik} to set
 an upper limit on the field strength at small scales. The method
 is based on the fact that the dissipation of small scale
 magnetic field can alter the chemical potential of the CMB. The
 current upper limit on the chemical potential \cite{fixsen}, yields
 an upper limit on the magnetic field, $B < 3 \times 10^{-8} {\rm G}$
 at the comoving scale of $\sim 400 {\rm pc}$. It is amazing that 
 this limit is so close to the lower limit for $B$ in Eq. (21),
 for the cyclotron emission to be efficient. The implication is 
 that if Eq. (21) is satisfied, cyclotron emission suppresses 
 any chemical potential, while if Eq. (21) is not satisfied, 
 the induced chemical potential is less than the observational
 limit. Therefore, both cases are consistent with the observations.

  The theoretical estimate for the magnitude of $\beta_M$, generated 
 in the QCD phase transition  at $T_c \sim 150 {\rm MeV}$, which relies 
 on the presence of hydrodynamic instabilities produced by expanding
 bubble walls, is about
 $10^{-29}$ at a $10 {\rm Mpc}$ comoving scale\cite{sigl}.
 This translates to $\beta_M \sim 10^{-8}$ at $1 {\rm pc}$, the 
 comoving Hubble radius at the phase transition. This value
 is barely enough to satisfy Eq. (20), at $T = 1 {\rm keV}$,
 where double-Compton ceases to be efficient but falls below
 the the required limit for lower temperatures. 
 
   On the other hand, another way of generating magnetic 
 field is via Electroweak
 phase transition \cite{baym}. The maximum value of $\beta_M$ generated 
 in this case can be as large as $10^{-3}$ which does satisfy
 Eq.(20).

  In summary, we introduce cyclotron emission as a potential 
  thermalization process of the CMB spectrum.
  We found that the process of relaxation
 of the chemical potential, in a regime that Comptonization is efficient, 
 is linear instead of exponential for 
 large deviations of the chemical potential from zero. As a result, 
 the relaxation time is
 proportional to the original value of the chemical potential.
  Requiring that the plasma effects do not suppress this process sets
 a lower limit on the magnitude of the magnetic field. While observational
 constraints are not conclusive about if the primordial magnetic
 field is larger than this lower limit or not, there are theoretical 
 estimates that suggest magnetic fields large enough for 
 cyclotron thermalization process to be efficient, are produced 
 in the early universe.

   I would like to thank Bruce Draine for useful discussions,
 and David Spergel for his supervision on this work.

\end{document}